\documentclass[12pt
  ]
{article}

\begin{document}

\title{Property of $cn$-$\bar n\bar c$ System in $\tilde U(12)$-Scheme and
$X(3872)/Y(3940)$}


\author{Muneyuki Ishida\\
  Department of Physics, Faculty of Sciences and Engineering,\\
  Meisei University, Hino, Tokyo 191-8506, Japan}

\date{}

\maketitle

\begin{center}
{\bf abstract}
\end{center}
\begin{small}
The properties of four quark $cn$-$\bar n\bar c$ states
are investigated as $cn$ di-quark and $\bar n \bar c$ di-antiquark system
in $\tilde U(12)$-classification scheme of hadrons.
We consider the negative-parity di-quark and di-antiquark in ground states,
and the properties of $X(3872)$ and $Y(3940)$ are consistent, respectively, with those 
of the $J^{PC}=1^{++}$ and $2^{++}$ states from these negative-parity di-quark and di-antiquark.
Their narrow-widths are explained from an orthogonality of spinor wave functions.
The properties of ground-state $cs$-$\bar s \bar c$ system are also predicted in this scheme.    
\end{small}


\section{Introduction}
\label{sec1}

The discovery of $X(3872)$\cite{X} and $Y(3940)$\cite{Y} by Belle presented new problems on 
hadron spectroscopy. Because of their decay-properties,
they are considered to be non-$c\bar c$ states.
We consider the four-quark explanation\cite{polosa} of $X(3872)/Y(3940)$, where 
the relevant particles are considered as the $cn$ di-quark and $\bar n \bar c$ di-antiquark system.
This explanation is criticized from the observed decay widths.
The $cn$-$\bar n \bar c$ system decays to $D\bar D^{(*)}$ and $J/\psi\rho$(or $J/\psi\omega)$.
They are so-called OZI super-allowed processes, 
and proceed without no $q\bar q$-pair production or annihilation. 
Their widths are expected to be so large as a few hundred MeV or more.
On the other hand the observed widths of $X(3872)$ and $Y(3940)$ are very small, 
$\Gamma_X < 2.3$MeV\cite{X} and $\Gamma_Y=87\pm 22\pm 26$MeV\cite{Y}, respectively.

A few years ago\cite{[1]}, we have proposed a covariant level-classification scheme of hadrons
based on $\tilde U(12)$ spin-flavor group. It is a relativistic generalization of the 
$SU(6)_{SF}$-scheme in non-relativistic quark model(NRQM), and the squared-mass spectra of hadrons including 
light constituent quarks are classified as the representations of $\tilde U(12)$\cite{[2]}.
Each spinor index corresponding to light constituent quark in composite hadron WF is expanded by 
free Dirac spinors $u_{r,s}(v_\mu )$ called urciton spinors. At the rest frame 
$u_{+,s}({\bf 0} )$ has the upper two-component which corresponds to the Pauli spinor 
appearing in NRQM, while the $u_{-,s}({\bf 0} )$ has the lower two-component
representing the relativistic effect.
The $r$ index of $u_{r,s}$ represents the eigen-value of $\rho_3$, and the corresponding freedom
is called $\rho$-spin, while the ordinary Pauli-spin freedom is called $\sigma$-spin.
In the $\tilde U(12)$-scheme, 
the $u_{-,s}$ is supposed to appear as new degrees of freedom for light constituents,
being independent of $u_{+,s}$,
and we predict the existence of a number of extra states out of the $SU(6)_{SF}$-framework
for meson and baryon systems, called chiral states.

The heavy-light $Q\bar q$ mesons in ground states are classified as ${\bf 12}^*$, 
which includes the chiral $J^P=0^+/1^+$ states with positive-parity 
as well as the negative-parity $0^-/1^-$ states\cite{[2]}.
The properties of $D_s(2317)/D_s(2460)$ are naturally explained as
the chiral states in $c\bar s$ system.
Applying the $\tilde U(12)$-scheme,
the heavy-light $Qq$ di-quark forms {\bf 12}
in ground states, which includes chiral $J^P=0^-/1^-$ states with negative parity as well as the 
positive parity $0^+/1^+$ states.

There is a kind of phenomenological conservation rule applicable to chiral states, 
called $\rho_3$-line rule\cite{baryon,SIshida}.
In transitions of chiral states, the $\rho_3$ along each spectator-quark line is 
approximately conserved. 

There is an interesting possibility that the observed $X(3872)$ and $Y(3940)$ are chiral states,
which are made from negative-parity $cn$ di-quark and $\bar n\bar c$ di-antiquark, 
and that their narrow-width properties are explained by 
the $\rho_3$-line rule.
In this talk we investigate this possibility,
whether the $X(3872)$ and $Y(3940)$ are explained as $cn$-$\bar n\bar c$ states.

\section{Di-quark and Di-antiquark Spinor Wave Function}

The negative-parity $cn$ di-quark($\bar n\bar c$ di-antiquark), called chiral state, 
is denoted as $\Gamma_{cn}^\chi (\Gamma_{\bar n \bar c}^\chi )$, which has
negative $\rho_3^n(\rho_3^{\bar n})$, while the ordinary
positive-parity di-quark(di-antiquark), called Pauli state, is denoted as 
$\Gamma_{cn}^P(\Gamma_{\bar n \bar c}^P)$,
which is composed of consitituents both with positive $\rho_3$.
Their explicit forms are given in ref.\cite{MI}. 

The $\Gamma_{cn}$ 
is related with the WF of $c\bar n$-meson $H_{c\bar n}$ in heavy quark effective theory, as
$\Gamma_{cn} = H_{c\bar n} C^\dagger$. 
The WF of positive(negative)-parity di-quark corresponds to the WF of negative(positive)-parity meson.
The positive-parity mesons, which are chiral states, become heavier\cite{[2]} than the 
negative-parity Pauli mesons through spontaneous chiral symmetry breaking($\chi$SB),
which is described by the Yukawa coupling to the scalar $\sigma$-meson nonet 
$s=s^i\lambda^i/\sqrt 2$ in the framework of $SU(3)$ linear $\sigma$ model\cite{LsM}.
In $Qq$ di-quark systems, because of the $C^\dagger$ factor, the chiral states 
are expected to have smaller masses than the Pauli states. 
The mass shift $\delta M^{\chi ,P}$ in $\chi$SB is given by
\begin{eqnarray}
\delta M^{\chi ,P} &=& g_\sigma s_0\ 
           \langle \Gamma_{cn}^{\chi ,P} \overline{\Gamma_{cn}^{\chi ,P}} \rangle
     = g_\sigma s_0\ 
           \langle H_{c\bar n}^{\chi ,P} C^\dagger C^\dagger 
                   \overline{H_{c\bar n}^{\chi ,P}} \rangle
     = - g_\sigma s_0\ \langle H_{c\bar n}^{\chi ,P} \overline{H_{c\bar n}^{\chi ,P}} \rangle ,
\ \ \ \ \ \ \ \ 
\label{eq1}
\end{eqnarray}
where $\overline{\Gamma_{cn}^{\chi ,P}}\equiv \gamma_4 (\Gamma_{cn}^{\chi ,P})^\dagger \gamma_4$, and 
the change of sign comes from $(C^\dagger)^2=-1$.
The $s_0$ is vacuum expectation value of $s$ and $g_\sigma$ is the Yukawa coupling constant.
The chiral mass splitting $\Delta M_\chi (\equiv \delta M^P-\delta M^\chi ) =2g_\sigma s_0$ 
is predicted with 242MeV\cite{[2]}.   

Concerning $cn$-$\bar n\bar c$ system, 
we expect three combinations of $\Gamma_{cn}$-$\Gamma_{\bar n\bar c}$:
The $\Gamma_{cn}^P$-$\Gamma_{\bar n\bar c}^P$ denoted as Pauli-Pauli states,
the $\Gamma_{cn}^P$-$\Gamma_{\bar n\bar c}^\chi$
(or $\Gamma_{cn}^\chi$-$\Gamma_{\bar n\bar c}^P$) denoted as Pauli-chiral(or chiral-Pauli) states, 
and the $\Gamma_{cn}^\chi$-$\Gamma_{\bar n\bar c}^\chi$ denoted as chiral-chiral states.
Among them the chiral-chiral states have the smallest masses, the chiral-Pauli states are heavier
by $\Delta M_\chi$, and the Pauli-Pauli states are further heavier by $\Delta M_\chi$.
The Pauli-Pauli(Pauli-chiral) states decay to Pauli-chiral(chiral-chiral) states 
through pion emission in $S$-wave, and the widths are expected to be so large as a GeV.
Accordingly, the Pauli-Pauli states and Pauli-chiral states are not observed
as resonances, but as non-resonant backgrounds.
We focus our interests on chiral-chiral states in the following.

\section{Oscillator WF and Ground-State Mass Spectra}

We consider spin-spin interaction to 
predict the mass spectra of ground states. We use   
the space WF in joint spring quark model(JSQM)\cite{JSQM},
where the $cn$ di-quark ($\bar n\bar c$ di-antiquark) is in color ${\bf 3}^*({\bf 3})$,
and they are connected by spring potential.
The space-coordinates of $c$, $n$, $\bar n$ and $\bar c$ are denoted, respectively, as 
${\bf x}_1$, ${\bf x}_2$, ${\bf x}_3$ and ${\bf x}_4$, and the 
eigen-modes are numerically given by\cite{MI}
\begin{eqnarray} 
{\bf s} &=& ({\bf x}_1-{\bf x}_2-{\bf x}_3+{\bf x}_4)/\sqrt 2,\ \ 
{\bf R}  = 0.785 ({\bf x}_1-{\bf x}_4) + 0.083 ({\bf x}_2-{\bf x}_3),\nonumber \\ 
{\bf U} &=& 0.365 ({\bf x}_1-{\bf x}_4) - 0.862 ({\bf x}_2-{\bf x}_3).\ \ 
\label{eq4}
\end{eqnarray}
We may regard {\bf s} as relative coordinate in di-(anti)quark,
{\bf R}({\bf U}) approximately as relative coordinate between $c$ and $\bar c$($n$ and $\bar n$). 
The internal space WF is given by
\begin{eqnarray}
f_s({\bf s}) f_R({\bf R}) f_U({\bf U}) &=& 
\left( \frac{\beta_s\beta_R\beta_U}{\pi^3} \right)^{\frac{3}{4}} e^{-\frac{\beta_s}{2}{\bf s}^2}
e^{-\frac{\beta_R}{2}{\bf R}^2}e^{-\frac{\beta_U}{2}{\bf U}^2},
\label{eq5}
\end{eqnarray}
where $\beta_s=0.129$GeV$^2$, $\beta_R=0.254$GeV$^2$ and $\beta_U=0.115$GeV$^2$.

By using the WF (\ref{eq5}), we can estimate the spin-spin interaction,
$
H_{SS} \propto -\frac{1}{m_i m_j} \delta^{(3)} ({\bf x}_i-{\bf x}_j) 
\mbox{\boldmath $\sigma^{(i)}$}\cdot \mbox{\boldmath $\sigma^{(j)}$}$.
The proprtional constant is fixed by using
$m_{D^*}-m_D$=142MeV as input, and
the mass-shifts 
$\Delta M_{SS}$ in ground-states of $cn$-$\bar n\bar c$ system are given in Table \ref{tab3}.
The mass of $X(3872)$, which plausibly $J^{PC}$=$1^{++}$\cite{Yabsley},
 is used as input to fix overall mass-shift.
\begin{table}
\begin{tabular}{l|cc|c|ccc}
type   & $J^{PC}$ & $\Delta M_{SS}$ & Mass & \multicolumn{3}{c}{decay channels}\\ 
\hline
$\Gamma_{cn}^{0^-}$-$\Gamma_{\bar n\bar c}^{0^-}$ & $0^{++}$ & -65 & 3824 
            & $(J/\psi \rho )$, & $\eta_c \pi$, & $D\bar D$\\
$\frac{1}{\sqrt 2}( \Gamma_{cn}^{0^-}$-$\Gamma_{\bar n\bar c}^{1^-}
                  + \Gamma_{cn}^{1^-}$-$\Gamma_{\bar n\bar c}^{0^-} )$
                                                  & $1^{+-}$ & -27 & 3862
            & $J/\psi\pi$, & $\eta_c\rho$ & \\
$\frac{1}{\sqrt 2}( \Gamma_{cn}^{0^-}$-$\Gamma_{\bar n\bar c}^{1^-}
                  - \Gamma_{cn}^{1^-}$-$\Gamma_{\bar n\bar c}^{0^-} )$
                                                  & $1^{++}$ & -17 & \underline{3872}
            & $J/\psi\rho$ &  & $D^0\bar D^{0*}$ \\
$\Gamma_{cn}^{1^-}$-$\Gamma_{\bar n\bar c}^{1^-}$ & $0^{++}$ & -22 & 3867 
            & $(J/\psi\rho )$, & $\eta_c\pi$, & $D\bar D$\\
$\Gamma_{cn}^{1^-}$-$\Gamma_{\bar n\bar c}^{1^-}$ & $1^{+-}$ &   0.& 3889
            & $J/\psi\pi$, & $\eta_c\rho$, & $D\bar D^*$\\
$\Gamma_{cn}^{1^-}$-$\Gamma_{\bar n\bar c}^{1^-}$ & $2^{++}$ & +43 & 3932
            & $J/\psi\rho$, & & $D\bar D^{(*)}$(D-wave)\\
\hline
\end{tabular}
\caption{Masses of ground-state $cn$-$\bar n\bar c$ system (in MeV). 
Decay channels of $I$=1 states are shown. For $I$=0, $\rho (\pi )$
is replaced with $\omega (\eta )$.
}
\label{tab3}
\end{table}
The $2^{++}$ state
is predicted with mass 3932MeV, which is consistent with the experimental value
of $Y(3940)$, 
$M=3943\pm 11 \pm 13$MeV\cite{Y}.
At the same time, the existence of the other 4 states
with $J^{PC}=0^{++}$ and $1^{+-}$ is predicted in the same mass region.

\section{Strong Decay Widths}

({\it Overlapping of spinor WF})\ \ \ 
The decay mechanism of the relevant quark-rearrangement processes is unknown, however,
it may be reasonable to assume that the decay matrix elements are proportional to the
overlappings of the initial and final WFs. 
The overlapping of spinor WFs, denoted as $A_s$, are given for the decays to 
$c\bar n+n\bar c$ and $c\bar c+n\bar n$, respectively, by 
\begin{eqnarray}
A_s (c\bar n + n\bar c) &=& \langle \Gamma_{\bar n\bar c}^\chi 
     \overline{H}_{\bar D}(v_{\bar D})(-iv_{\bar D}\cdot\gamma)\
    {}^t\Gamma_{cn}^\chi\  {}^t[iv_D\cdot\gamma \overline{H}_D(v_D) ] \rangle\nonumber\\
  &=& \langle \Gamma_{\bar n\bar c}^\chi \ 
    B(v_{\bar D}) \overline{H}_{\bar D}  \bar B(v_{\bar D})\
    {}^t\Gamma_{cn}^\chi\  {}^t[ B(v_D) \overline{H}_D \bar B(v_D) ] \rangle
 \nonumber\\
A_s(c\bar c + n\bar n) &=& \langle \Gamma_{\bar n\bar c}^\chi 
    \overline{H}_{\psi}(v_\psi)\Gamma_{cn}^\chi\ 
    {}^t[iv_\rho\cdot\gamma \overline{H}_\rho (v_\rho) (-iv_\rho\cdot\gamma) ] \rangle \nonumber\\
 &=& \langle \Gamma_{\bar n\bar c}^\chi \ 
    B(v_\psi)\overline{H}_{\psi}\bar B(v_\psi)\ \Gamma_{cn}^\chi\ 
    {}^t[B(v_\rho)\overline{H}_\rho  \bar B(v_\rho) ] \rangle ,
\label{eq10}
\end{eqnarray}
where $D\bar D$ and $J/\psi \rho$ are taken as examples of 
$c\bar n+n\bar c$ and $c\bar c + n\bar n$.
The $\overline{H}(v)$ is the spinor WF of the final meson with velocity $v_\mu (\equiv P_\mu /M)$, 
and $\overline{H}$ is the one at the rest frame with velocity $v_{0\mu}=(0,0,0;i)$. 
They are related by Lorentz booster 
$B(v),\bar B(v)=ch\theta \pm \rho_1\mbox{\boldmath $n$}\cdot\mbox{\boldmath $\sigma$}sh\theta$ as
$\overline{H}(v)=B(v)\overline{H}\bar B(v)$, where the $ch\theta ,sh\theta =\sqrt{\frac{E\pm M}{2M}}$.
In Eq.~(\ref{eq10}), the change of $\rho_3$ on each quark-line in the decay
occurs only through $\rho_1$ term
in $B(v)$ and $\bar B(v)$, which is proportional to $sh\theta$. It is related with the momentum 
${\bf P}$ of the final meson as $ch\theta sh\theta =\frac{\bf P}{2M}$, which is  
much smaller than unity for the relevant decays because of their small phase space. 
Thus, the $\rho_3$ quantum number along every the spectator-quark line is approximately 
conserved\cite{baryon}.

In the chiral-chiral states, the $n$ in $cn$ di-quark and 
the $\bar n$ in $\bar n\bar c$ di-antiquark are in negative $\rho_3$ state. Thus,  
their decay amplitudes to the Pauli mesons, of which $\rho_3$ of constituents are all positive, 
are doubly suppressed with the factor $(\frac{\bf P}{2M})^2$.
In $\tilde U(12)$-scheme,
the WF of vector mesons, $J/\psi$, $\rho$, $\omega$, $D^*$ and $\bar D^*$, are 
commonly 
given by $\overline{H}=(1+iv_0\cdot\gamma)\frac{i\epsilon\cdot\gamma}{2\sqrt 2}$, where
the $\rho_3$ of constituents are all positive, $(\rho_3^q,\rho_3^{\bar q})=(+,+)$. 
On the other hand, the WF of pseudoscalar mesons, $\pi$ and $\eta$, which are Nambu-Goldstone bosons,
are $\overline{H}=\frac{i\gamma_5}{2}$, which has doubly-negative $\rho_3$ component as 
$(\rho_3^q,\rho_3^{\bar q})=\frac{1}{\sqrt 2}(\ (+,+)+(-,-)\ )$(, while the WF of $D$ and $\eta_c$ is 
$\overline{H}=(1+iv_0\cdot\gamma)\frac{\gamma_5}{2\sqrt 2}$, which has  
$(\rho_3^c,\rho_3^{\bar c})=(+,+)$). 
Thus, in the relevant decays of the chiral-chiral states,
the transition amplitudes to $J/\psi \rho (\omega )$, $\eta_c \rho (\omega)$ and $D\bar D^{(*)}$
are strongly suppressed from $\rho_3$-line rule, while the ones to $J/\psi \pi (\eta)$
and $\eta_c \pi (\eta)$ are not suppressed. 
The decay widths of the latter channels are expected to be a few hundred MeV or a GeV.

Thus, when the initial states have no $\pi (\eta)$ decay modes, they have narrow
widths, and are expected to be observed as resonant particles. 
The possible decay channels of the ground six states are shown in Table \ref{tab3}.
{\it Among them, only two states with $J^{PC}=1^{++}$ and $2^{++}$ have no $\pi (\eta)$ decay modes
and are considered to be detected as resonances.} 
%
This theoretical expectation is consistent with the present experimetnal situation:
Only two states, $X(3872)$ and $Y(3940)$, are observed as the exotic candidates in the
relevant mass region.

({\it Decay widths})\ \ \ 
We can also calculate the overlappings of space WFs, denoted as $\tilde F_x$. 
Their numerical values are $\tilde F_x(c\bar n + n\bar c)=0.96$GeV$^{-\frac{3}{2}}$
and $\tilde F_x(c\bar c + n\bar n)=1.05$GeV$^{-\frac{3}{2}}$, respectively.
We neglect the difference between them, and use a common value $\tilde F_x\simeq 1$GeV$^{-\frac{3}{2}}$.

The width $\Gamma$ of the decay $X$(or $Y$)$\rightarrow A+B$ is given by 
\begin{eqnarray}
\Gamma &=&  a |{\bf P}|  ( E_A E_B M \tilde F_x^2 )  | A_s |^2
\label{eq9}
\end{eqnarray}
with one dimension-less parameter $a$, where $M$ is the mass of initial 4 quark state, and 
$E_A(E_B)$ is the energy of the final meson $A(B)$. 
The $|{\bf P}|$ is the momentum of final meson.
The 
widths for the
 relevant decay modes of $X(3872)$ and $Y(3940)$ are given in 
Table \ref{tab4}. The parameter $a$ is fixed with $a=18.8$, which is determined by the 
the total width of $Y(3940)$, $\Gamma_Y=87$MeV, as input.  
\begin{table}
\begin{tabular}{c|c|c|c||c|c|c}
$R$ & $J/\psi\omega$ & $D\bar D$ & $D\bar D^*$ &  &  $J/\psi\rho$ & $J/\psi\omega$ \\
\hline
$\Gamma_Y(R)$ & 60.3\ cos$^2\theta_Y$ & 20.1 & 6.6 &
 $\Gamma_X(R)$ & (1.2 sin$^2\theta_X$) & (0.2 cos$^2\theta_X$)\\
\hline
\end{tabular}
\caption{Decay widths of $Y(3940)$, $\Gamma_Y$, and $X(3872)$, $\Gamma_X$ in MeV. 
For the $\Gamma_Y(D\bar D^*)$, $\bar D D^*$ is also included. 
The cos$\theta_{X,Y}$(sin$\theta_{X,Y}$) are $I$=0(1) component of the flavor WFs of $X,Y$.
See ref.\cite{MI} for details.
}
\label{tab4}
\end{table}
The partial decay rates of $Y(3940)$
are fairly consistent with the present experiments. 
By taking the $I$=0 component cos$\theta_Y \simeq 1$, the main decay mode of $Y(3940)$ is considered
to be $J/\psi\omega$.
Experimentally the $Y(3940)$ is observed in $J/\psi\omega$, not in $D\bar D$ and $D\bar D^*$.

\section{Concluding Remarks}

The $X(3872)$ and $Y(3940)$ have the properties of 
the four quark $cn$-$\bar n\bar c$ states with $J^{PC}=1^{++}$ and $2^{++}$, respectively,
which come from $0^-$-$1^-$ and $1^-$-$1^-$ combinations of 
$cn$ chiral di-quark and $\bar n\bar c$ chiral di-antiquark in $\tilde U(12)$-scheme.
Their mass difference is consistently explained by the spin-spin interaction between constituents
in JSQM.
Their narrow widths are explained from a phenomenological selection rule, 
coming from an orthogonality of spinor WFs, called $\rho_3$-line rule.
Only these two states are expected to be observed as resonant particles in ground states, and the 
other states have very large widths and are not observed as resonances but as non-resonant backgrounds.
\begin{table}
\begin{tabular}{c|c|c|c||c|c}
   & $\Gamma_Y (D_s\bar D_s)$ & $\Gamma_Y (D_s\bar D_s^*)$ & $\Gamma_Y(J/\psi\phi)$ & 
   &  $\Gamma_X^{\rm tot}$ \\
\hline
$Y_{cs-\bar s\bar c}^{2^{++}}$(4082) & 23MeV & 8MeV & 11MeV &
$X_{cs-\bar s\bar c}^{1^{++}}$(4152) & $<1$MeV  \\
\hline
\end{tabular}
\caption{The masses and widths for the ground $cs$-$\bar s\bar c$ states 
with $J^{PC}=1^{++}$, $X_{cs-\bar s\bar c}^{1^{++}}$ , and $2^{++}$, $Y_{cs-\bar s\bar c}^{2^{++}}$. 
%
The total width of $Y_{cs-\bar s\bar c}^{2^{++}}$ is estimated by the sum as 
$\Gamma^{\rm tot}_{Y^{2^{++}}_{cs-\bar s\bar c}}$=23+8+11=42MeV.
 }
\label{tab5}
\end{table}

In order to check this interpretation,
we present the predictions for the properties of ground $cs$-$\bar s\bar c$ states in Table \ref{tab5}. 
Their masses are estimated simply by adding $2(M_{D_s^*}-M_{D^{*0}})\simeq 210$MeV
to the corresponding $cn$-$\bar n\bar c$ states. 
Their decay widths are estimated by using Eq.~(\ref{eq9}) 
with the common parameter $a$.
The mass of $X_{cs-\bar s\bar c}^{1^{++}}$ is expected to be quite close
to the $D_s\bar D_s^*$ threshold and its decay is strongly suppressed.
The $J/\psi \phi$ is un-open channel. Thus, the main decay mode is considered to be 
electromagnetic ones of order keV.
The main decay mode of $Y_{cs\bar s\bar c}^{2^{++}}$ is $D_s\bar D_s$. 
In order to check the four quark nature of this state, 
it is necessary to observe $J/\psi\phi$ decay.\\
{\it The author would like to express his sincere gratitude to Prof. S. Ishida and Dr. T. Maeda
for useful information and comments. He also acknowledges deeply Professor M.~Oka for 
helpful comments and suggestions.}

\end{document}